\documentclass[%
 reprint,
 amsmath,amssymb,
 aps,
]{revtex4-1}

\usepackage{graphicx}
\usepackage{dcolumn}
\usepackage{bm}

\begin{document}

\preprint{APS/123-QED}

\title{Kinetic theory of overpopulated gluon systems with inelastic processes}

\author{Zhengyu Chen$^{1}$}
\affiliation{$^1$ Department of Physics, Tsinghua University, Beijing 100084, China}

\date{\today}

\begin{abstract}
In this work, the role of inelastic processes in the formation of a transient Bose-Einstein condensation  (BEC) is investigated based on kinetic theory. We calculate the condensation rate for an overpopulated gluon system which is assumed to be in thermal equilibrium and with the presence of a BEC. The matrix elements of the inelastic processes are chosen as the isotropic one and the gluons are considered to have a finite mass. Our calculations indicate that the inelastic processes can hinder the formation of a BEC since the negatively infinite net condensation rate can destroy any BEC instantly.
\end{abstract}

\pacs{Valid PACS appear here}
\maketitle


\section{Introduction}\label{sec:introduction}

The deconfined matter produced in experiments of ultrarelativistic heavy-ion collisions, the
quark-gluon plasma (QGP), has been widely considered as a nearly perfect fluid close to local thermal equilibrium. Meanwhile, the mechanism of the fast thermalization in the early stage still demands further investigations. The early stage of heavy-ion collisions can be well described within the Color-Glass-Condensate (CGC) effective field theory\cite{McLerran:1993ni,McLerran:1993ka,McLerran:1994vd}. After collision of two nuclei, gluons are freed and evolve to form a so-called glasma \cite{Gelis:2010nm,Lappi:2006fp,Weigert:2005us} through a very short isotropization stage \cite{Gelis:2013rba,Kurkela:2015qoa}. Gluons at this time are far from thermal equilibrium and their number density in such glasma can be overwhelmingly higher than its corresponding thermal equilibrium density with the same energy density. On the way of thermalization the excess of gluons might be removed into a Bose-Einstein condensate (BEC). 

A BEC is the macroscopic occupation in the ground state due to the fundamental consequence of quantum statistics. In the framework that only binary elastic collisions considered \cite{Blaizot:2011xf,Berges:2012us,Blaizot:2013lga,Xu:2014ega,Meistrenko:2015mda,Epelbaum:2015vxa,Zhou:2017zql,Chen:2018mwr}, the nonequilibrium dynamics of the BECs formation have been investigated within either kinetic approach or classical field theory. Whether a BEC can be formed (grow) is still undebate if inelastic processes are considered. The authors in \cite{Huang:2013lia} find that, compared with the purely elastic case, the inelastic processes can catalyze the onset of dynamical BEC to occur faster. Obtaining a simple kinetic equation that allows for an analytic description of the most important small momentum regimes, the analysis in \cite{Blaizot:2016iir} suggests that the formation of a BEC is strongly hindered by the inelastic processes. Based on the two-particle irreducible (2PI) formalism to next-to-leading order in the $1/N$ expansion, the study in \cite{Tsutsui:2017uzd} show that the formation of a BEC is hindered by particle number changing processes. And recently, within a kinetic approach by including interactions of massive bosons with constant and isotropic cross sections, the simulations in \cite{Lenkiewicz:2019glw} demonstrate that BECs are highly unlikely if inelastic collisions are significantly participating in the dynamical gluonic evolution. Different from the above mentioned works focusing on the dynamical formation of a transient BEC, we calclutate the net rate of the condensation ( net condensation rate ) for an overoccupied gluon system in thermal equilibrium, which can in turn give a hint whether a BEC can be formed.

The net condensation rate, which is the counterbalance of the production and evaporation processes, can determine the growth, duration and decay of a BEC. The net condensation rate can be zero, negative or positive depending on the exact stage the system has evolved to and the microscopic interactions among the bosons. Assuming a boson system in thermal equilibrium and with the presence of a BEC, the net condensation rate can be calculated within kinetic theory. A zero or positive net condensation rate indicates that the BEC can exactly exist on the condition it can be formed. If the negative net condensation rate is finite, a formed BEC can durate for a finite short time, while the negatively infinite net rate indicates that the BEC can not be formed at all since any formed BEC will decay instantly. As mentioned in our former work \cite{Chen:2018mwr}, according to a direct estimation from the equilibrium distribution functions involved in the collision terms, number-changing processes cannot destroy a massless gluon BEC, while they do for a massive gluon BEC and the decay rate depends on the exact form of the matrix elements. 

In this work, we give a detailed calculation of the condensation rates for an overpopulated gluon system in thermal equilibrium, and the gluons are considered to have a finite mass. The interactions among gluons include both elastic and inelastic processes and for matrix elements we employ the isotropic one. We mention that our framework is almost the same as that of Ref.~\cite{Lenkiewicz:2019glw}, and we calculate the condensation rate at equilibrium by an analytical way while the authors in Ref.~\cite{Lenkiewicz:2019glw} study the time evolution of the condensate from a far from equilibrium glasma-type initial condition by numerically solving the integro-differential Boltzmann equation.  The rest of the paper is organized as follows. We present the kinetic equation for the Bose-Einstein condensation in Sec.~\ref{sec:kinetic_equation}. The net condensation rates are derived for isotropic matrix element in Sec.~\ref{sec:rate_of_conden}. We summarize in Sec.~\ref{sec:summary}. Details on the derivations of rate equations are given in Appendices.

\section{Kinetic equation for Bose-Einstein condensation}\label{sec:kinetic_equation}
The particle distribution function $f(p,t)$ in the presence of a BEC is decomposed into two parts $f=f^g+f^c$, where $f^g$ and $f^c$ denote the distribution of gas (noncondensate) and condensate particles, respectively. We consider an isotropic and homogeneous gluon system. The gluons are considered to have finite mass $m$, while the generalization of the following calculation for massless particles is
straightforward. Thus, the momentum distribution in thermal equilibrium has the form
\begin{equation}
f_{eq}({\bf p})=\frac{1}{e^{(E-m)/T}-1}+(2\pi)^3n_c\delta^{(3)}({\bf p}) \,, 
\label{eq:f_eq}
\end{equation}
where $T$ is the temperature, $E=\sqrt{{\bf p}^2+m^2}$ is the energy of the on-shell particles and $n_c$ the condensate particle density with zero momentum. The Boltzmann equation considering both elastic and inelastic processes takes the following form:
\begin{widetext}
\begin{eqnarray}
\frac{\partial f_1}{\partial t} &=& \frac{1}{2E_1}\int d{\rm \Gamma_2}\frac{1}{2!}\int d{\rm \Gamma_3}d{\rm \Gamma_4}\vert M_{34\rightarrow 12}\vert ^2 [f_3f_4(1+f_1)(1+f_2)-f_1f_2(1+f_3)(1+f_4)]\nonumber \\
&&\times(2\pi)^4\delta^{(4)}(p_3+p_4-p_1-p_2) \nonumber \\
&&+\frac{1}{2E_1}\int d{\rm \Gamma_2}\frac{1}{3!}\int d{\rm \Gamma_3}d{\rm \Gamma_4}d{\rm \Gamma_5}\vert M_{345\rightarrow 12}\vert ^2 [f_3f_4f_5(1+f_1)(1+f_2)-f_1f_2(1+f_3)(1+f_4)(1+f_5)] \nonumber \\
&&\times(2\pi)^4\delta^{(4)}(p_3+p_4+p_5-p_1-p_2) \nonumber \\
&&+\frac{1}{2!}\frac{1}{2E_1}\int d{\rm \Gamma_2}d{\rm \Gamma_3}\frac{1}{2!}\int d{\rm \Gamma_4}d{\rm \Gamma_5}\vert M_{45\rightarrow 123}\vert ^2 [f_4f_5(1+f_1)(1+f_2)(1+f_3)-f_1f_2f_3(1+f_4)(1+f_5)] \nonumber \\
&&\times(2\pi)^4\delta^{(4)}(p_4+p_5-p_1-p_2-p_3)\;,
\label{eq:Boltzmann_eq}
\end{eqnarray}
\end{widetext} 
where $f_i=f_i({\bf r},{\bf p}_i,t)$ and $d{\rm \Gamma}_i=d^3p_i/(2E_i)/(2\pi)^3$, $i=1,2,3,4,5$. The matrix elements $\vert M \vert ^2$ determine the microscopic interactions among particles, which are chosen as both the isotropic one and the exact one based on pQCD in this work. The collision terms on the right-hand side of Eq.~(\ref{eq:Boltzmann_eq}) take into account quantum statistics via Bose enhancement factors $(1+f_i)$, which can lead to the correct long-time equilibrium solution for bosons. The first term accounts for elastic processses, the second term the inelastic processes with the observed particle $p_1$ on the two-particle side and the third term the inelastic processes with the observed particle $p_1$ on the three-particle side. The factorial numbers denote the multiple counting of the identical boson particles. 

Denoting gas particles by $g$ and condensate particles by $c$, the elastic collision processes $g+g\longleftrightarrow g+g$ and $g+c\longleftrightarrow g+g$ are considered in Ref.~\cite{Zhou:2017zql}. The inelastic processes contain $g+g\longleftrightarrow g+g+g$, $g+c\longleftrightarrow g+g+g$, $g+g+c\longleftrightarrow g+g$ and $g+c+c\longleftrightarrow g+g$. Thus, Eq.~(\ref{eq:Boltzmann_eq}) can be separated into two parts corresponding to gas and condensate particles. Since we foucus on the rates of the condensation in this work, we thus give the Boltzmann equation for the condensate particles 
\begin{widetext}
\begin{eqnarray}
\frac{\partial f^c_1}{\partial t} &=& \frac{1}{2E_1}\int d{\rm \Gamma_2}\frac{1}{2!}\int d{\rm \Gamma_3}d{\rm \Gamma_4}\vert M_{gg\rightarrow gc}\vert ^2 [f^g_3f^g_4f^c_1(1+f_2)-f^c_1f^g_2(1+f^g_3)(1+f^g_4)]\nonumber \\
&&\times(2\pi)^4\delta^{(4)}(p_3+p_4-p_1-p_2) \nonumber \\
&&+\frac{1}{2E_1}\int d{\rm \Gamma_2}\frac{1}{3!}\int d{\rm \Gamma_3}d{\rm \Gamma_4}d{\rm \Gamma_5}\vert M_{ggg\rightarrow gc}\vert ^2 [f^g_3f^g_4f^g_5f^c_1(1+f^g_2)-f^c_1f^g_2(1+f^g_3)(1+f^g_4)(1+f^g_5)] \nonumber \\
&&\times(2\pi)^4\delta^{(4)}(p_3+p_4+p_5-p_1-p_2) \nonumber \\
&&+\frac{1}{2!}\frac{1}{2E_1}\int d{\rm \Gamma_2}d{\rm \Gamma_3}\frac{1}{2!}\int d{\rm \Gamma_4}d{\rm \Gamma_5}\vert M_{gg\rightarrow ggc}\vert ^2 [f^g_4f^g_5f^c_1(1+f^g_2)(1+f^g_3)-f^c_1f^g_2f^g_3(1+f^g_4)(1+f^g_5)] \nonumber \\
&&\times(2\pi)^4\delta^{(4)}(p_4+p_5-p_1-p_2-p_3) \nonumber \\
&&+\frac{1}{2!}\frac{1}{2E_1}\int d{\rm \Gamma_2}d{\rm \Gamma_3}\frac{1}{2!}\int d{\rm \Gamma_4}d{\rm \Gamma_5}\vert M_{gg\rightarrow gcc}\vert ^2 [f^g_4f^g_5f^c_1f^c_2(1+f^g_3)-f^c_1f^c_2f^g_3(1+f^g_4)(1+f^g_5)] \nonumber \\
&&\times(2\pi)^4\delta^{(4)}(p_4+p_5-p_1-p_2-p_3) \;.
\label{eq:CondBoltzmann_eq}
\end{eqnarray}
\end{widetext} 

Integrating the above Boltzmann equation for the condensate particles with their momenta gives the time derivative of the density, or in other words, the condensation rate. The contribution from elastic processes is denoted by $R_{el}$, and from the three inelastic processes involving condensation are denoted by $R_{gc}$, $R_{ggc}$, $R_{gcc}$, respectively. The rate equations are given as follows:
\begin{widetext}
\begin{eqnarray}
\label{eq:R_el}
R_{el} &=& \frac{1}{2}\int d{\rm \Gamma_1}d{\rm \Gamma_2} \int d{\rm \Gamma_3}d{\rm \Gamma_4}\vert M_{gg\rightarrow gc}\vert ^2 [f^g_3f^g_4f^c_1(1+f_2)-f^c_1f^g_2(1+f^g_3)(1+f^g_4)]\nonumber \\
&&\times(2\pi)^4\delta^{(4)}(p_3+p_4-p_1-p_2) \;, \\
\label{eq:R_gc}
R_{gc} &=& \frac{1}{6}\int d{\rm \Gamma_1}d{\rm \Gamma_2} \int d{\rm \Gamma_3}d{\rm \Gamma_4}d{\rm \Gamma_5}\vert M_{ggg\rightarrow gc}\vert ^2 [f^g_3f^g_4f^g_5f^c_1(1+f^g_2)-f^c_1f^g_2(1+f^g_3)(1+f^g_4)(1+f^g_5)] \nonumber \\
&&\times(2\pi)^4\delta^{(4)}(p_3+p_4+p_5-p_1-p_2) \;, \\
\label{eq:R_ggc}
R_{ggc} &=& \frac{1}{4}\int d{\rm \Gamma_1}d{\rm \Gamma_2}d{\rm \Gamma_3} \int d{\rm \Gamma_4}d{\rm \Gamma_5}\vert M_{gg\rightarrow ggc}\vert ^2 [f^g_4f^g_5f^c_1(1+f^g_2)(1+f^g_3)-f^c_1f^g_2f^g_3(1+f^g_4)(1+f^g_5)] \nonumber \\
&&\times(2\pi)^4\delta^{(4)}(p_4+p_5-p_1-p_2-p_3) \;, \\
\label{eq:R_gcc}
R_{gcc} &=& \frac{1}{4}\int d{\rm \Gamma_1}d{\rm \Gamma_2}d{\rm \Gamma_3} \int d{\rm \Gamma_4}d{\rm \Gamma_5}\vert M_{gg\rightarrow gcc}\vert ^2 [f^g_4f^g_5f^c_1f^c_2(1+f^g_3)-f^c_1f^c_2f^g_3(1+f^g_4)(1+f^g_5)] \nonumber \\
&&\times(2\pi)^4\delta^{(4)}(p_4+p_5-p_1-p_2-p_3)\;.
\end{eqnarray}
\end{widetext}    

We can first take a detailed look at the terms involving the distribution functions on the right hand sides of the above rate equations. Considering the system in thermal equilibrium, $f^g_i({\bf p}_i)=1/(e^{(E_i-m)/T}-1)$ and $f^c_i=(2\pi)^3n_c\delta^{(3)}({\bf p}_i)$, we have   
\begin{widetext}
\begin{eqnarray}
f^g_3f^g_4f^c_1(1+f_2)-f^c_1f^g_2(1+f^g_3)(1+f^g_4) &=& 0 \;, \\
f^g_3f^g_4f^g_5f^c_1 (1+f^g_2)-f^c_1f^g_2(1+f^g_3)(1+f^g_4)(1+f^g_5) &=& f^c_1f^g_2f^g_3f^g_4f^g_5e^{(E-3m)/T}(e^{m/T}-1)\;, \\
f^g_4f^g_5f^c_1(1+f^g_2)(1+f^g_3) -f^c_1f^g_2f^g_3(1+f^g_4)(1+f^g_5) &=& f^c_1f^g_2f^g_3f^g_4f^g_5e^{(E-3m)/T}(1-e^{m/T})\;, \\
f^g_4f^g_5f^c_1f^c_2 (1+f^g_3)-f^c_1f^c_2f^g_3(1+f^g_4)(1+f^g_5) &=& f^c_1f^c_2f^g_4f^g_5(1+f^g_3)(1-e^{m/T})\;,
\end{eqnarray}
\end{widetext} 
where $E$ is the total energy of the processes. The net condensation rate of the elastic processes is always zero as they do not destroy the BEC. For the three inelastic processes, we can see that the net condensation rates are all zero if $m=0$, which means the inelastic processes can not destroy the BEC of a massless boson system. For a massive boson system, the net rate for the processes $g+c\longleftrightarrow g+g+g$ is positive, while the other two processes, $g+g+c\longleftrightarrow g+g$ and $g+c+c\longleftrightarrow g+g$, are negative. How these three processes counterbalance depends on the microscopic interactions between particles, and the total net condensation rate determine whether the BEC can be formed. It needs a more detailed investigation for the massive case. The rate equations can be symplified by integrating out the delta functions. Once the specific form of the matrix elements is given, which accounts for the microscopic interactions among particles, we can get the net condensation rate of the corresponding processes.

\section{Rate of the condensation}\label{sec:rate_of_conden}
The rate equation for the elastic processes has been investigted in Ref.~\cite{Zhou:2017zql} for a gluon system, which has the form
\begin{eqnarray}
R_{el} &=& \frac{n_c}{64\pi^3}\int d{E_3}d{E_4}  [f^g_3f^g_4-f^g_2(1+f^g_3+f^g_4)] \nonumber \\
&& \times E \left[ \frac{\vert M_{34\rightarrow 12}\vert ^2}{s}\right]_{s=2mE}  \;,
\label{eq:R_el1}
\end{eqnarray}
where $E=E_3+E_4$ is the total energy, $P=\vert {\bf p}_3+ {\bf p}_4 \vert$ is the total momentum, and $s=E^2-{\bf P}^2$ is the invariant mass. $m$ denotes the particle mass at rest. The authors draw the conclusion for a massless gluon system that, in order to describe the condensation with a finite rate, the ratio $\vert M_{34\rightarrow 12}\vert ^2/s$ at $s=0$ should be nonzero and finite. Both the isotropic matrix element and the pQCD based one calculated by using the Hard-Thermal-Loop (HTL) treatment, which are adopted in Ref.~\cite{Zhou:2017zql}, fulfill this requirement. At equilibrium, the net condensation rate is zero, while the scattering rates with a condensate particle in the processes, $g+c\longleftrightarrow g+g$, are infinite (the difference of the two infinite rates is zero). Therefore, in these two cases, the condensation is expected as demonstrated by the dynamical simulation, if the gluons are initially overoccupied and only binary elastic processes are considered. 

After a lengthy calculation for the integrals of the right-hand side of Eq.~(\ref{eq:R_gc}) $\sim$ Eq.~(\ref{eq:R_gcc}), which details are given in Appendix~\ref{app:rate_eq}, we obtain
\begin{widetext}
\begin{eqnarray}
\label{eq:R_gc1}
R_{gc} &=& \frac{n_c}{3\cdot 2^{10}\cdot \pi^5} \int dE_3 dE_4 dE_5 \left\lbrace f^g_3f^g_4f^g_5 (1+f^g_2)-f^g_2(1+f^g_3)(1+f^g_4)(1+f^g_5)\right\rbrace  \nonumber \\
&& \times \left[ p_3+p_4-\vert p_3-p_4 \vert\right] E \left[ \frac{\vert M_{345\rightarrow 12}\vert ^2}{s} \right]_{s=2mE}    \;, \\
\label{eq:R_ggc1}
R_{ggc} &=& \frac{n_c}{64\pi }\int \frac{d^3p_4}{(2\pi)^32E_4}\frac{d^3p_5}{(2\pi)^32E_5}  \int^{E^+_2}_{E^-_2} dE_2 \left\lbrace f^g_4f^g_5(1+f^g_2)(1+f^g_3) -f^g_2f^g_3(1+f^g_4)(1+f^g_5) \right\rbrace   \nonumber \\
&& \times \frac{1}{P} \left[ \frac{\vert M_{45\rightarrow 123}\vert ^2}{m} \right]_{E-E_2-\sqrt{({\bf P}-{\bf p}_2)^2+m^2}=m}  \,, \\
\label{eq:R_gcc1}
R_{gcc} &=& \frac{n^2_c}{512\pi^3}\int dE_4 dE_5  \left[ f^g_4f^g_5 (1+f^g_3)-f^g_3(1+f^g_4)(1+f^g_5)\right] (4E-3m)^2 \left[ \frac{\vert M_{45\rightarrow 123}\vert ^2}{s^2}\right]_{s=4mE-3m^2} \;,
\end{eqnarray} 
\end{widetext}
where the integration limits for $dE_2$ in Eq.~(\ref{eq:R_ggc1}) hve the form
$E^{\pm}_2 = \frac{1}{2} \left[ E-m \pm P\sqrt{1-\frac{4m^2}{s-2mE+m^2}} \, \right]$.

According to the above expressions for the condensation rates, we can give the following analysis for massless gluons. For the processes, $g+c+c\longleftrightarrow g+g$, the momenta of all gas particles should be parallel, which is the same as in elastic scatterings. In order to describe the condensation of massless gluons with a finite net rate, $\vert M \vert ^2/s^2$ should be nonzero and finite at $s=0$. Meanwhile, the scattering rates with a condensate particle are infinite at equilibrium. For the processes, $g+c\longleftrightarrow g+g+g$, the net condensation rate is proportional to $\vert M \vert ^2/s$ at $s=0$. Since $\vert M \vert ^2/s^2$ at $s=0$ is nonzero and finite, $\vert M \vert ^2/s=0$ at $s=0$. This means that processes $g+c\longleftrightarrow g+g+g$  will not play any role in the condensation. For the processes, $g+g+c\longleftrightarrow g+g$, momenta of gas particles are not necessarily parallel, processes can occur at any $s$, and the scattering rates are finite.

Similar to the judgement in last section, it is also straightforward to draw the conclusion that the net condensation rate for massive gluons is negative since the term $f^g_4f^g_5 (1+f^g_3)-f^g_3(1+f^g_4)(1+f^g_5)$ appeared in Eq.~(\ref{eq:R_gcc1}) (the processes $g+c+c\longleftrightarrow g+g$ are the dominant inelastic processes) is negative in this case. Whether this negative net rate is finite or infinite, which determine the transient BEC formation, depends on the form of the matrix elements for the inelastic processes. The tree-level matrix element for the process $g+g\rightarrow g+g+g$ is given by evaluating 25 Feynman diagrams \cite{Berends:1981rb}, and due to its complexity, a widely used approximation to this leading order perturbative QCD (pQCD) matrix element is Gunion-Bertsch (GB) formula \cite{Gunion:1981qs}, which was originally derived for gluon emission from quark-quark scattering \cite{Gunion:1981qs} and later explicitly used to derive the soft gluon emission from gluon-gluon scattering \cite{Biro:1993qt,Wong:1996ta}. The GB formula, which is involved with the transverse momentum of the radiated gluon and the transverse exchanged momentum, has a relatively simpler expression compared with the leading order pQCD matrix element. This simplified formula for soft gluon emission has been widely used to investigate the phenomena emerged in relativistic heavy-ion collisions. However, there appear some discrepancies between the results by employing the GB formula and the exact leading order pQCD matrix element, and several works \cite{Das:2010hs, Abir:2010kc, Bhattacharyya:2011vy, Fochler:2013epa} have made corrections or improvements for GB formula. When considering realistic gluon systems, even the exact leading order pQCD matrix element is not enough, adding loop contributions or summation over higher orders are needed. In this work, we employ the isotropic matrix element just like the case in Ref.~\cite{Lenkiewicz:2019glw} and leave the investigation on the realistic gluon systems with QCD interactions for the next job. In the following, giving the gluons a finite and small mass, we calculate the net rates for the three inelastic processes when the system is in thermal equilibrium.

If the scattering is isotropic with constant cross section, the squared matrix elements for the inelastic $2\leftrightarrow 3$ processes can be expressed as \cite{El:2012cr}:
\begin{eqnarray}
\label{eq:con_matrix}
\vert M_{2\rightarrow 3}\vert ^2 &=& 192\pi^3 \sigma_{23} \;, \\
\label{eq:con_matrix1}
\vert M_{3\rightarrow 2}\vert ^2 &=& \vert M_{2\rightarrow 3}\vert ^2 /d_g \;,
\end{eqnarray}
where $\sigma_{23}$ is the constant total cross section and $d_g$ is the gluon degeneracy factor. Applying this isotropic matrix element to Eq.~(\ref{eq:R_gc1}) $\sim$ Eq.~(\ref{eq:R_gcc1}), which details are given in Appendix~\ref{app:rate_eq_cim}, we can get the rates at equilibrium, 
\begin{widetext}
\begin{eqnarray}
\label{eq:R_gc2}
R_{gc} &=& \frac{n_c T^4}{3\cdot 2^{10}\cdot \pi^5} (e^{-m/T}-e^{-2m/T}) \frac{\vert M_{345\rightarrow 12} \vert ^2}{m} \Big\{ \int^{e^{-m/T}}_{0} dx_3 \int^{e^{-m/T}}_{x_3} dx_4 \frac{\sqrt{\ln^2 x_4-(m/T)^2}}{(e^{-m/T}-x_3)(e^{-m/T}-x_4)(e^{-m/T}-x_3x_4)}   \nonumber \\  
&& \times \left[ \ln (e^{-m/T}-x_3x_4)-\ln (0) \right]  + e^{m/T}\int^{e^{-m/T}}_{0} dx_3 \frac{\sqrt{\ln^2 x_3-(m/T)^2}}{(e^{-m/T}-x_3)(1-x_3)} \nonumber \\
&& \times \left[ \ln \frac{e^{-m/T}-x^2_3}{e^{-m/T}-x_3} \ln (0) +\frac{1}{2}\ln^2(e^{-m/T}-x^2_3)-\frac{1}{2}\frac{m^2}{T^2} +\int^{x_3}_{0} dx_4 \frac{\ln (e^{-m/T}-x_3x_4)}{e^{-m/T}-x_4} \right] \Big\} \;, \\
\label{eq:R_ggc2}
R_{ggc} &=& \frac{n_cT}{64\cdot (2\pi)^5 } (1-e^{m/T}) \frac{\vert M_{45\rightarrow 123}\vert ^2}{m} \int dp_4dp_5 \int d\cos \theta \frac{p^2_4p^2_5}{E_4E_5} \frac{1}{P} \frac{1}{(e^{(E_4-m)/T}-1)(e^{(E_5-m)/T}-1)}    \nonumber \\
&& \times \frac{1}{1-e^{(3m-E_4-E_5)/T}} \ln \frac{\cosh \frac{E-3m+P\sqrt{1-\frac{4m^2}{s-2mE+m^2}}}{2T}-1}{\cosh \frac{E-3m-P\sqrt{1-\frac{4m^2}{s-2mE+m^2}}}{2T}-1} \Theta (E-3m) \Theta ((E-m)^2-{\bf P}^2-4m^2)  \,, \\
\label{eq:R_gcc2}
R_{gcc} &=& \frac{n^2_c T^2}{512\pi^3} \frac{\vert M_{45\rightarrow 123} \vert ^2}{m^2} \Big\{ -2\ln^2 (0) -[11\frac{m}{T}-4ln(e^{m/T}-1)] \ln (0) -4\frac{m^2}{T^2} +6\frac{m}{T}ln(e^{m/T}-1) -\frac{1}{2} ln^2(e^{m/T}-1)  \nonumber \\
&& -\frac{\pi^2}{6} +Li_2(\frac{1}{1-m/T}) +Li_2(e^{m/T}) +Li_2(e^{-m/T}) \Big\}   \;, 
\end{eqnarray} 
\end{widetext}
where $\theta$ in Eq.~(\ref{eq:R_ggc2}) denotes the angle between ${\bf p}_4$ and ${\bf p}_5$, and the last three terms in Eq.~(\ref{eq:R_gcc2}) denote the polylogarithm functions $Li_n(z)$ with $n=2$ and $z=1/(1-m/T), e^{m/T}$ and $e^{-m/T}$, respectively.

For a massive gluon system, Eq.~(\ref{eq:R_gc2}) and Eq.~(\ref{eq:R_ggc2}) can only be calculated numerically by giving a particular ratio of the finite mass over temperature $m/T$, for example, letting $m = 100$ MeV and $T = 400$ MeV which may correspond to an overpopulated gluon system in ultrarelativistic heavy-ion collisions. The quantity $0$ appeared in Eq.~(\ref{eq:R_gc2}) and Eq.~(\ref{eq:R_gcc2}) denotes the limit converge to $0^+$, for example, $ln(0)$ means $\lim \limits_{x \to 0^+} ln(x)$. The rates will have the forms: $R_{gc} \sim B_1 ln(0) + C_1$, $R_{ggc} \sim C_2$ and $R_{gcc} \sim A_3 ln^2(0) + B_3 ln(0) + C_3$, where $A_i, B_i, C_i$ represent finite constants ($B_1$, $C_2$ and $A_3$ are negative). Therefore, the condensation rate of the processes $g+c\longleftrightarrow g+g+g$ is positively infinite, the processes $g+g+c\longleftrightarrow g+g$ is negatively finite and the processes $g+c+c\longleftrightarrow g+g$ is negatively infinite. Summing all the rates in Eq.~(\ref{eq:R_gc2}) $\sim$ Eq.~(\ref{eq:R_gcc2}), we obtain the net rate $\sim A ln^2(0) + B ln(0) + C$, where the coefficients $A$ is negative, $B$ may be either positive or negative depending on $m, T, n_c $ and $\vert M_{45\rightarrow 123}\vert ^2$. The net rate is negatively infinite since $ln^2(0)$ is the higher order infinity compared with $ln(0)$. These indicate that the formation of a BEC is hindered by the inelastic processes with the isotropic matrix element since the negatively infinite net rate for the condensation can destroy any BEC instantly.

The investigations \cite{Blaizot:2000fc,Andersen:2002ey,Caron-Huot:2007rwy} of the thermodynamics based on resummed perturbation theory indicate that the most important plasma effects are the thermal masses acquired by the hard thermal particles, thus the gluon thermal mass or Debye mass can be treated as regulator for the infrared divergence. At first sight, the negatively infinite rate is reminiscent of the infrared singularities, like the infrared divergence in soft bremsstrahlung. However, it is not the fact in our considered case, since the gluons are given a finite constant mass which can cure that divergence just like the thermal mass or Debye mass. The physical origin of the infinite terms appeared in Eq.~(\ref{eq:R_gc2}) and Eq.~(\ref{eq:R_gcc2}) is due to the fact that the gluon system is in thermal equilibrium fulfilling Bose-Einstein distribution and the Bose enhancement effect denoted by the Bose factors. We have checked that the condensation rates would not be infinite if the gluon system is not in thermal equilibrium, or even in thermal equilibrium but with Boltzmann distribution. 

\section{Summary}\label{sec:summary}

In this paper, we investigate the role of inelastic processes in the formation of a transient Bose-Einstein condensation based on kinetic theory. Different from most of the existing works focusing on the dynamical formation of a transient BEC from a far from equilibrium glasma-type initial condition, we calclutate the net condensation rate for an overoccupied gluon system in thermal equilibrium and with the presence of a BEC. The inelastic processes cannot destroy a massless gluon BEC, while they do for a massive gluon BEC and the decay rate depends on the exact form of the matrix elements. The matrix elements of the inelastic processes are chosen as the isotropic one and the gluons are considered to have a finite mass. Our results show that the formation of a BEC is hindered by the inelastic processes since the negatively infinite net rate for the condensation can destroy any BEC instantly. Thus, it is likely that the formation of Bose-Einstein condensates in the very early stage of  ultrarelativistic heavy-ion collisions is hindered by the inelastic processes. Our results confirm the work of Ref.~\cite{Lenkiewicz:2019glw} by an analytical way since we implement the same isotropic matrix elements, the same interaction processes and the gluons are taken to be massive.    

It should be mentioned that the present work is limited to an isotropic and homogeneous gluon system, the case of a longitudinally expanding system which is more closely related to heavy-ion collisions needs to be considered. The role of expansion in the BEC formation \cite{Epelbaum:2015vxa,Berges:2015ixa} will be our following subject.  

\begin{acknowledgments}
Z.C. would like to thank Yan Zhu for helpful discussions and appreciate Zhe Xu for enlightening comments and suggestions. This work was financially supported by the National Natural Science Foundation of China under Grants No. 11890710, No. 11890712, and No. 12035006. 
\end{acknowledgments}

\appendix

\begin{widetext}
\section{\label{app:rate_eq}Rate equations of the condensation}

In this appendix, we derive the rate equations, Eq.~(\ref{eq:R_gc}) $\sim$ Eq.~(\ref{eq:R_gcc}), to the simplified form, Eq.~(\ref{eq:R_gc1}) $\sim$ Eq.~(\ref{eq:R_gcc1}).

\subsection{\label{appA:gc}Rate for $g+c\longleftrightarrow g+g+g$ }
Eq.~(\ref{eq:R_gc}) for the prcocess $g+c\longleftrightarrow g+g+g$ consists of the condensation part and evaporation part, which can be named as $R^{gain}_{gc}$ and $R^{loss}_{gc}$. In the following we carry out integrations in $R^{gain}_{gc}$ explicitly. 

At first we integrate over $d^3p_2$ with help of the delta function $\delta^{(3)}({\bf p}_3+{\bf p}_4+{\bf p}_5-{\bf p}_1-{\bf p}_2)$ and obtain
\begin{eqnarray}
\label{eq:APP_R_gc}
R^{gain}_{gc} &=& \frac{1}{6}\int \frac{d^3p_3}{(2\pi)^32E_3}\frac{d^3p_4}{(2\pi)^32E_4}\frac{d^3p_5}{(2\pi)^32E_5}\frac{d^3p_1}{(2\pi)^32E_1} \frac{1}{2(E-E_1)} \vert M_{345\rightarrow 12}\vert ^2 2\pi \delta[F({\bf p}_1)] f^g_3f^g_4f^g_5f^c_1(1+f^g_2) \;,
\end{eqnarray}
where $f^g_2=f^g(E-E_1,{\bf P}-{\bf p}_1;t)$, $E=E_1+E_2=E_3+E_4+E_5$ is the total energy and ${\bf P}={\bf p}_1+{\bf p}_2={\bf p}_3+{\bf p}_4+{\bf p}_5$ is the total momentum. $\delta[F({\bf p}_1)]$ indicates the energy conservation, where
\begin{eqnarray}
F({\bf p}_1) &=& E-E_1-E_2 = E-E_1-\sqrt{({\bf P}-{\bf p}_1)^2+m^2} \;.
\end{eqnarray}

Using the identity
\begin{equation}
\label{eq:measure}
\int dE_1d^3p_1\delta(E_1^2-p_1^2-m^2)= \int \frac{d^3p_1}{2E_1} \;.
\end{equation}
and $f^c_1=(2\pi)^3n_c\delta^{(3)}({\bf p}_1)$ we then rewrite Eq.~(\ref{eq:APP_R_gc}) to
\begin{eqnarray}
R^{gain}_{gc} &=& \frac{\pi n_c}{3}\int \frac{d^3p_3}{(2\pi)^32E_3}\frac{d^3p_4}{(2\pi)^32E_4}\frac{d^3p_5}{(2\pi)^32E_5} f^g_3f^g_4f^g_5 \int dE_1d^3p_1 \frac{1}{2(E-E_1)} \vert M_{345\rightarrow 12}\vert ^2 (1+f^g_2) \nonumber \\
&& \times \delta[F({\bf p}_1)] \delta(E_1^2-p_1^2-m^2) \delta^{(3)}({\bf p}_1) \;,
\end{eqnarray}

As the next we integrate over $d^3p_1$ and then $dE_1$ using the delta function $\delta^{(3)}({\bf p}_1)$ and $\delta[F({\bf p}_1)]$
\begin{eqnarray}
\label{eq:APP_Rgain_gc1}
R^{gain}_{gc} &=& \frac{\pi n_c}{3}\int \frac{d^3p_3}{(2\pi)^32E_3}\frac{d^3p_4}{(2\pi)^32E_4}\frac{d^3p_5}{(2\pi)^32E_5} f^g_3f^g_4f^g_5 \int dE_1 \frac{1}{2(E-m)} \vert M_{345\rightarrow 12}\vert ^2 (1+f^g_2) \delta(E_1^2-m^2) \nonumber \\
&& \times \delta(E-m-\sqrt{P^2+m^2}) \nonumber \\
&=& \frac{\pi n_c}{3}\int \frac{d^3p_3}{(2\pi)^32E_3}\frac{d^3p_4}{(2\pi)^32E_4}\frac{d^3p_5}{(2\pi)^32E_5} f^g_3f^g_4f^g_5 (1+f^g_2) \frac{1}{2m} \frac{1}{2(E-m)} \vert M_{345\rightarrow 12}\vert ^2  \nonumber \\
&& \times \delta(E-m-\sqrt{P^2+m^2}) \;,
\end{eqnarray}
where $f^g_2=f^g(\sqrt{P^2+m^2},{\bf P};t)$. 

Without loss of generality, we choose ${\bf p}_3$ to along z-axis, and the total momentum $P$ has the form 
\begin{eqnarray}
\label{eq:APP_Rgain_gc2}
P &=& \vert {\bf p}_3+{\bf p}_4+{\bf p}_5 \vert \nonumber \\
&=&\left\lbrace p_3^2+p_4^2+p_5^2+2p_3p_4\cos\theta_4 +2p_3p_5\cos\theta_5 +2p_4p_5\left[ \sin\theta_4\sin\theta_5\cos(\phi_4-\phi_5) +\cos\theta_4\cos\theta_5 \right] \right\rbrace ^{1/2} \,.
\end{eqnarray}

We can integrate Eq.~(\ref{eq:APP_Rgain_gc2}) over the solid angles of ${\bf p}_3$ and ${\bf p}_4$,
\begin{eqnarray}
\label{eq:APP_Rgain_gc3}
R^{gain}_{gc} &=& \frac{\pi n_c}{3} \int 4\pi \frac{p^2_3dp_3}{(2\pi)^32E_3} \int 2\pi \frac{p^2_4dp_4\sin\theta_4 d\theta_4}{(2\pi)^32E_4} \int \frac{p^2_5dp_5\sin\theta_5 d\theta_5 d\phi_5}{(2\pi)^32E_5} f^g_3f^g_4f^g_5 (1+f^g_2) \frac{1}{2m} \frac{1}{2(E-m)} \vert M_{345\rightarrow 12}\vert ^2  \nonumber \\
&& \times \delta(E-m-\sqrt{P^2+m^2}) \;,
\end{eqnarray} 

The integral over $\cos\phi_5$ can be carried out using the delta function and gives
\begin{eqnarray}
\label{eq:APP_Rgain_gc4}
R^{gain}_{gc} &=& \frac{n_c}{3\cdot 2^9\cdot \pi^6} \int \frac{p^2_3dp_3}{E_3} \int \frac{p_4dp_4 d\theta_4}{E_4} \int \frac{p_5dp_5 d\theta_5}{E_5} f^g_3f^g_4f^g_5 (1+f^g_2) \frac{1}{4m}  \vert M_{345\rightarrow 12}\vert ^2 \frac{1}{\sin {\bar \phi_5}} \;,
\end{eqnarray} 
where $\sin {\bar \phi_5}$ is the solution of $E-m-\sqrt{P^2+m^2}=0$, and has the form 
\begin{eqnarray}
\sin {\bar \phi_5} &=&  \left\lbrace  4p^2_4p^2_5\sin^2 \theta_4 \sin^2 \theta_5 -(E^2-2mE-p_3^2-p_4^2-p_5^2-2p_3p_4\cos\theta_4 -2p_3p_5\cos\theta_5 -2p_4p_5\cos\theta_4 \cos\theta_5)^2 \right\rbrace ^{1/2}  \nonumber \\
&& \times \frac{1}{2p_4p_5\sin\theta_4 \sin\theta_5} \;.
\end{eqnarray} 
Thus, we have
\begin{eqnarray}
\label{eq:APP_Rgain_gc5}
R^{gain}_{gc} &=& \frac{n_c}{3\cdot 2^8\cdot \pi^6} \int \frac{p^2_3dp_3}{E_3} \int \frac{p^2_4dp_4 }{E_4} \int \frac{p^2_5dp_5}{E_5} \int^{u^+_4}_{u^-_4} du_4 \int^{u^+_5}_{u^-_5} du_5 f^g_3f^g_4f^g_5 (1+f^g_2) \frac{1}{4m}  \vert M_{345\rightarrow 12}\vert ^2  \nonumber \\
&& \times \left\lbrace -Au^2_4+Bu_4+C \right\rbrace ^{-1/2} \;,
\end{eqnarray} 
where $u_4=\cos \theta_4$, $u_5=\cos \theta_5$, $A=4p^2_3p^2_4+4p^2_4p^2_5+8p_3p^2_4p_5u_5$, $B=4(E^2-2mE-p_3^2-p_4^2-p_5^2-2p_3p_5u_5)(p_3p_4+p_4p_5u_5)$ and $C=4p^2_4p^2_5(1-u^2_5)-(E^2-2mE-p_3^2-p_4^2-p_5^2-2p_3p_5u_5)^2$. According to the requirement $\sin {\bar \phi_5}\leq 1$, we can obtain the integration limits for $u_4$ 
\begin{equation}
u^{\pm}_4=\frac{-B\pm \sqrt{B^2+4AC}}{-2A} \;.
\end{equation}

Carry out the integral over $u_4$ and then $u_5$, we get
\begin{eqnarray}
\label{eq:APP_Rgain_gc6}
R^{gain}_{gc} &=& \frac{n_c}{3\cdot 2^8\cdot \pi^5} \int \frac{p^2_3dp_3}{E_3} \int \frac{p^2_4dp_4 }{E_4} \int \frac{p^2_5dp_5}{E_5} f^g_3f^g_4f^g_5 (1+f^g_2) \frac{1}{4m}  \vert M_{345\rightarrow 12}\vert ^2 \frac{1}{2p_4} \int^{1}_{-1} du_5 \nonumber \\
&& \times \frac{1}{\sqrt{p^2_3+p^2_5+2p_3p_5u_5}} \nonumber \\
&=& \frac{n_c}{3\cdot 2^9\cdot \pi^5} \int \frac{p_3dp_3}{E_3} \int \frac{p_4dp_4 }{E_4} \int \frac{p_5dp_5}{E_5} f^g_3f^g_4f^g_5 (1+f^g_2) \frac{1}{4m}  \vert M_{345\rightarrow 12}\vert ^2 \left[  p_3+p_5-\vert p_3-p_5 \vert \right]  \nonumber \\
&=& \frac{n_c}{3\cdot 2^9\cdot \pi^5} \int dE_3 dE_4 dE_5 f^g_3f^g_4f^g_5 (1+f^g_2) \frac{1}{4m}  \vert M_{345\rightarrow 12}\vert ^2 \left[ p_3+p_4-\vert p_3-p_4 \vert\right]  \;,
\end{eqnarray} 
in the last equality we have applied the symmetry of ${\bf p}_3$, ${\bf p}_4$ and ${\bf p}_5$.

The integrals in $R^{loss}_{gc}$ proceed similarly as those shown above. We obtain $R^{loss}_{gc}$ by replacing $f^g_3f^g_4f^g_5(1+f^g_2)$ in Eq.~(\ref{eq:APP_Rgain_gc6}) with $f^g_2(1+f^g_3)(1+f^g_4)(1+f^g_5)$. Since $s=E^2-{\bf P}^2$, the constraint $E-m-\sqrt{P^2+m^2}=0$ is equivalent to $s=2mE$. We have finally 
\begin{eqnarray}
\label{eq:APP_Rgc}
R_{gc} &=& \frac{n_c}{3\cdot 2^{10}\cdot \pi^5} \int dE_3 dE_4 dE_5 \left\lbrace f^g_3f^g_4f^g_5 (1+f^g_2)-f^g_2(1+f^g_3)(1+f^g_4)(1+f^g_5)\right\rbrace  \nonumber \\
&& \times \left[ p_3+p_4-\vert p_3-p_4 \vert\right] E \left[ \frac{\vert M_{345\rightarrow 12}\vert ^2}{s} \right]_{s=2mE}    \;.
\end{eqnarray} 

\subsection{\label{appA:ggc}Rate for $g+g+c\longleftrightarrow g+g$ }
We carry out integrations in $R^{gain}_{ggc}$ explicitly, which has the form:
\begin{eqnarray}
\label{eq:APP_R_ggc1}
R^{gain}_{ggc} &=& \frac{1}{4}\int d{\rm \Gamma_1}d{\rm \Gamma_2}d{\rm \Gamma_3} \int d{\rm \Gamma_4}d{\rm \Gamma_5} \vert M_{45\rightarrow 123}\vert ^2 f^g_4f^g_5f^c_1(1+f^g_2)(1+f^g_3) (2\pi)^4 \delta^{(4)}(p_4+p_5-p_1-p_2-p_3) \;,
\end{eqnarray}
At first we integrate over $d^3p_3$ with help of the delta function $\delta^{(3)}({\bf p}_4+{\bf p}_5-{\bf p}_1-{\bf p}_2-{\bf p}_3)$ and obtain
\begin{eqnarray}
\label{eq:APP_R_ggc2}
R^{gain}_{ggc} &=& \frac{1}{4}\int \frac{d^3p_4}{(2\pi)^32E_4}\frac{d^3p_5}{(2\pi)^32E_5}\frac{d^3p_1}{(2\pi)^32E_1}\frac{d^3p_2}{(2\pi)^32E_2} \frac{1}{2(E-E_1-E_2)} \vert M_{45\rightarrow 123}\vert ^2 f^g_4f^g_5f^c_1(1+f^g_2)(1+f^g_3)  \nonumber \\
&&\times 2\pi \delta[F({\bf p}_1,{\bf p}_2)] \;,
\end{eqnarray}
where $f^g_3=f^g(E-E_1-E_2,{\bf P}-{\bf p}_1-{\bf p}_2;t)$, $E=E_1+E_2+E_3=E_4+E_5$ is the total energy and ${\bf P}={\bf p}_1+{\bf p}_2+{\bf p}_3={\bf p}_4+{\bf p}_5$ is the total momentum. $\delta[F({\bf p}_1,{\bf p}_2)]$ indicates the energy conservation, where
\begin{eqnarray}
F({\bf p}_1,{\bf p}_2) = E-E_1-E_2-E_3 = E-E_1-E_2-\sqrt{({\bf P}-{\bf p}_1-{\bf p}_2)^2+m^2} \;. \nonumber \\
\end{eqnarray} 
Using the identity Eq.~(\ref{eq:measure}) and $f^c_1=(2\pi)^3n_c\delta^{(3)}({\bf p}_1)$ we then rewrite Eq.~(\ref{eq:APP_R_ggc1}) to 
\begin{eqnarray}
\label{eq:APP_Rgain_ggc3}
R^{gain}_{ggc} &=& \frac{\pi n_c}{2}\int \frac{d^3p_4}{(2\pi)^32E_4}\frac{d^3p_5}{(2\pi)^32E_5} f^g_4f^g_5  \int \frac{d^3p_2}{(2\pi)^32E_2} dE_1 d^3p_1 \frac{1}{2(E-E_1-E_2)} \vert M_{45\rightarrow 123}\vert ^2 (1+f^g_2)(1+f^g_3) \nonumber \\
&&\times \delta^{(3)}({\bf p}_1) \delta(E_1^2-p_1^2-m^2) \delta[F({\bf p}_1,{\bf p}_2)] \;,
\end{eqnarray}
As the next we integrate over $d^3p_1$ and then $dE_1$ using the delta function $\delta^{(3)}({\bf p}_1)$ and $\delta[F({\bf p}_1,{\bf p}_2)]$
\begin{eqnarray}
\label{eq:APP_Rgain_ggc4}
R^{gain}_{ggc} &=& \frac{\pi n_c}{2}\int \frac{d^3p_4}{(2\pi)^32E_4}\frac{d^3p_5}{(2\pi)^32E_5} f^g_4f^g_5  \int \frac{d^3p_2}{(2\pi)^32E_2} dE_1 \frac{1}{2(E-E_1-E_2)} \vert M_{45\rightarrow 123}\vert ^2 (1+f^g_2)(1+f^g_3) \nonumber \\
&&\times \delta(E_1^2-m^2) \delta(E-E_1-E_2-\sqrt{({\bf P}-{\bf p}_2)^2+m^2}) \nonumber \\ 
&=& \frac{\pi n_c}{2}\int \frac{d^3p_4}{(2\pi)^32E_4}\frac{d^3p_5}{(2\pi)^32E_5} f^g_4f^g_5  \int \frac{d^3p_2}{(2\pi)^32E_2} \frac{1}{2\sqrt{({\bf P}-{\bf p}_2)^2+m^2}} \vert M_{45\rightarrow 123}\vert ^2 (1+f^g_2)(1+f^g_3) \nonumber \\
&&\times \delta((E-E_2-\sqrt{({\bf P}-{\bf p}_2)^2+m^2})^2-m^2) \;,
\end{eqnarray}

We carry out the integral over the solid angle of ${\bf p}_2$ by using the delta function and obtain
\begin{eqnarray}
\label{eq:APP_Rgain_ggc5}
R^{gain}_{ggc} &=& \frac{n_c}{32\pi }\int \frac{d^3p_4}{(2\pi)^32E_4}\frac{d^3p_5}{(2\pi)^32E_5} f^g_4f^g_5  \int \frac{p^2_2dp_2d\cos\theta_2}{E_2} \frac{1}{\sqrt{({\bf P}-{\bf p}_2)^2+m^2}} \vert M_{45\rightarrow 123}\vert ^2 (1+f^g_2)(1+f^g_3) \nonumber \\
&&\times \delta((E-E_2-\sqrt{({\bf P}-{\bf p}_2)^2+m^2})^2-m^2) \nonumber \\
&=& \frac{n_c}{32\pi }\int \frac{d^3p_4}{(2\pi)^32E_4}\frac{d^3p_5}{(2\pi)^32E_5} f^g_4f^g_5  \int \frac{p_2dp_2}{E_2} \frac{1}{P} \frac{1}{2m} \vert M_{45\rightarrow 123}\vert ^2 (1+f^g_2)(1+f^g_3) \nonumber \\
&=& \frac{n_c}{32\pi }\int \frac{d^3p_4}{(2\pi)^32E_4}\frac{d^3p_5}{(2\pi)^32E_5} f^g_4f^g_5  \int dE_2 \frac{1}{P} \frac{1}{2m} \vert M_{45\rightarrow 123}\vert ^2 (1+f^g_2)(1+f^g_3) \,,
\end{eqnarray}
the constrain $E-E_2-\sqrt{({\bf P}-{\bf p}_2)^2+m^2}=m$ is required for the above integration.

The integrals in $R^{loss}_{ggc}$ proceed similarly as those shown above. We obtain $R^{loss}_{ggc}$ by replacing $f^g_4f^g_5(1+f^g_2)(1+f^g_3)$ in Eq.~(\ref{eq:APP_Rgain_ggc5}) with $f^g_2f^g_3(1+f^g_4)(1+f^g_5)$. We have finally 
\begin{eqnarray}
\label{eq:APP_Rggc}
R_{ggc} &=& \frac{n_c}{64\pi }\int \frac{d^3p_4}{(2\pi)^32E_4}\frac{d^3p_5}{(2\pi)^32E_5}  \int^{E^+_2}_{E^-_2} dE_2 \left\lbrace f^g_4f^g_5(1+f^g_2)(1+f^g_3) -f^g_2f^g_3(1+f^g_4)(1+f^g_5) \right\rbrace   \nonumber \\
&& \times \frac{1}{P} \left[ \frac{\vert M_{45\rightarrow 123}\vert ^2}{m} \right]_{E-E_2-\sqrt{({\bf P}-{\bf p}_2)^2+m^2}=m}  \,,
\end{eqnarray}
the integration limits for $dE_2$ are obtained according to the constraint $E-E_2-\sqrt{({\bf P}-{\bf p}_2)^2+m^2}=m$ and $\vert \cos\theta_2 \vert \leq 1$,
\begin{equation}
E^{\pm}_2 = \frac{1}{2} \left[ E-m \pm P\sqrt{1-\frac{4m^2}{s-2mE+m^2}} \right] \;, 
\end{equation}
where $s=E^2-{\bf P}^2$.

\subsection{\label{appA:gcc}Rate for $g+c+c\longleftrightarrow g+g$}
We carry out integrations in $R^{gain}_{gcc}$ explicitly, which has the form:
\begin{eqnarray}
R_{gcc} &=& \frac{1}{4}\int d{\rm \Gamma_1}d{\rm \Gamma_2}d{\rm \Gamma_3} \int d{\rm \Gamma_4}d{\rm \Gamma_5}\vert M_{45\rightarrow 123}\vert ^2 f^g_4f^g_5f^c_1f^c_2(1+f^g_3) (2\pi)^4 \delta^{(4)}(p_4+p_5-p_1-p_2-p_3) \;,
\end{eqnarray}

At first we integrate over $d^3p_3$ with help of the delta function $\delta^{(3)}({\bf p}_4+{\bf p}_5-{\bf p}_1-{\bf p}_2-{\bf p}_3)$ and obtain
\begin{eqnarray}
\label{eq:APP_R_gcc}
R^{gain}_{gcc} &=& \frac{\pi}{2}\int \frac{d^3p_4}{(2\pi)^32E_4}\frac{d^3p_5}{(2\pi)^32E_5}\frac{d^3p_1}{(2\pi)^32E_1}\frac{d^3p_2}{(2\pi)^32E_2} \frac{1}{2(E-E_1-E_2)} \vert M_{45\rightarrow 123}\vert ^2 \nonumber \\
&& \times f^g_4f^g_5f^c_1f^c_2(1+f^g_3) \delta[F({\bf p}_1,{\bf p}_2)] \;,
\end{eqnarray}
where $f^g_3=f^g(E-E_1-E_2,{\bf P}-{\bf p}_1-{\bf p}_2;t)$, $E=E_1+E_2+E_3=E_4+E_5$ is the total energy and ${\bf P}={\bf p}_1+{\bf p}_2+{\bf p}_3={\bf p}_4+{\bf p}_5$ is the total momentum. $\delta[F({\bf p}_1,{\bf p}_2)]$ indicates the energy conservation, where
\begin{eqnarray}
F({\bf p}_1,{\bf p}_2) = E-E_1-E_2-E_3 = E-E_1-E_2-\sqrt{({\bf P}-{\bf p}_1-{\bf p}_2)^2+m^2} \;. \nonumber \\
\end{eqnarray} 
Using the identity Eq.~(\ref{eq:measure}) and $f^c_1=(2\pi)^3n_c\delta^{(3)}({\bf p}_1)$ we then rewrite Eq.~(\ref{eq:APP_R_gcc}) to 
\begin{eqnarray}
R^{gain}_{gcc} &=& \frac{\pi n_c}{2}\int \frac{d^3p_4}{(2\pi)^32E_4}\frac{d^3p_5}{(2\pi)^32E_5} f^g_4f^g_5  \int \frac{d^3p_2}{(2\pi)^32E_2} dE_1 d^3p_1 \frac{1}{2(E-E_1-E_2)} \vert M_{45\rightarrow 123}\vert ^2 f^c_2(1+f^g_3) \nonumber \\
&&\times \delta^{(3)}({\bf p}_1) \delta(E_1^2-p_1^2-m^2) \delta[F({\bf p}_1,{\bf p}_2)] \;,
\end{eqnarray}

As the next we integrate over $d^3p_1$ and then $dE_1$ using the delta function $\delta^{(3)}({\bf p}_1)$ and $\delta[F({\bf p}_1,{\bf p}_2)]$
\begin{eqnarray}
\label{eq:APP_Rgain_gcc1}
R^{gain}_{gcc} &=& \frac{\pi n_c}{2}\int \frac{d^3p_4}{(2\pi)^32E_4}\frac{d^3p_5}{(2\pi)^32E_5} f^g_4f^g_5  \int \frac{d^3p_2}{(2\pi)^32E_2} dE_1 \frac{1}{2(E-E_1-E_2)} \vert M_{45\rightarrow 123}\vert ^2 f^c_2(1+f^g_3) \nonumber \\
&&\times \delta(E_1^2-m^2) \delta(E-E_1-E_2-\sqrt{({\bf P}-{\bf p}_2)^2+m^2}) \nonumber \\ 
&=& \frac{\pi n_c}{2}\int \frac{d^3p_4}{(2\pi)^32E_4}\frac{d^3p_5}{(2\pi)^32E_5} f^g_4f^g_5  \int \frac{d^3p_2}{(2\pi)^32E_2} \frac{1}{2\sqrt{({\bf P}-{\bf p}_2)^2+m^2}} \vert M_{45\rightarrow 123}\vert ^2 f^c_2(1+f^g_3) \nonumber \\
&&\times \delta((E-E_2-\sqrt{({\bf P}-{\bf p}_2)^2+m^2})^2-m^2) \;.
\end{eqnarray}

Subsequently, we integrate over $d^3p_2$ using the delta function $\delta^{(3)}({\bf p}_2)$ and obtain
\begin{eqnarray}
\label{eq:APP_Rgain_gcc2}
R^{gain}_{gcc} &=& \frac{\pi n^2_c}{2}\int \frac{d^3p_4}{(2\pi)^32E_4}\frac{d^3p_5}{(2\pi)^32E_5} f^g_4f^g_5(1+f^g_3) \frac{1}{2m} \frac{1}{2\sqrt{{\bf P}^2+m^2}} \vert M_{45\rightarrow 123}\vert ^2   \delta((E-m-\sqrt{{\bf P}^2+m^2})^2-m^2) \;, \nonumber \\
\end{eqnarray}
where $f^g_3=f^g(\sqrt{{\bf P}^2+m^2}, {\bf P};t)$. We denote that $\theta$ is the angle between ${\bf p}_4$ and ${\bf p}_5$. Then we have
\begin{equation}
P=\vert {\bf p}_4+{\bf p}_5 \vert =\sqrt{p^2_4+p^2_5+2p_4p_5 {\rm cos}\theta}
\end{equation}

We assume that the distribution function $f^g_i$ is isotropic in momentum space. Therefore, $f^g_4=f^g(p_4,t)$, $f^g_5=f^g(p_5,t)$, $f^g_3=f^g(p,t)$, and we can integrate Eq.~(\ref{eq:APP_Rgain_gcc2}) over the solid angles of ${\bf p_4}$ and ${\bf p_5}$
\begin{eqnarray}
R^{gain}_{gcc} &=& \frac{\pi n^2_c}{256\pi^3}\int dp_4 dp_5 \frac{p^2_4p^2_5}{E_4E_5} f^g_4f^g_5 \int d \cos\theta \frac{1}{m} \frac{1}{\sqrt{{\bf P}^2+m^2}} \vert M_{45\rightarrow 123}\vert ^2 (1+f^g_3) 
\delta((E-m-\sqrt{{\bf P}^2+m^2})^2-m^2) \;, \nonumber \\
\end{eqnarray}

The integral over $\cos\theta$ can be carried out using the delta function and gives
\begin{eqnarray}
\label{eq:APP_Rgain_gcc3}
R^{gain}_{gcc} &=& \frac{n^2_c}{512\pi^3}\int dp_4 dp_5 \frac{p_4p_5}{E_4E_5} f^g_4f^g_5 (1+f^g_3)   \frac{1}{m^2} \left[ \vert M_{45\rightarrow 123}\vert ^2 \right]_{E-\sqrt{{\bf P}^2+m^2}=2m}   \nonumber \\
&=& \frac{n^2_c}{512\pi^3}\int dE_4 dE_5 f^g_4f^g_5 (1+f^g_3) \frac{1}{m^2} \left[ \vert M_{45\rightarrow 123}\vert ^2 \right]_{E-\sqrt{{\bf P}^2+m^2}=2m} \;,
\end{eqnarray}
The integrals in $R^{loss}_{gcc}$ proceed similarly as those shown above. We obtain $R^{loss}_{gcc}$ by replacing $f^g_4f^g_5(1+f^g_3)$ in Eq.~(\ref{eq:APP_Rgain_gcc3}) with $f^g_3(1+f^g_4)(1+f^g_5)$. Since $s=E^2-{\bf P}^2$, the constraint $E-\sqrt{P^2+m^2}=2m$ is equivalent to $s=m(4E-3m)$. We have finally 
\begin{eqnarray}
\label{eq:APP_Rgcc}
R_{gcc} &=& \frac{n^2_c}{512\pi^3}\int dE_4 dE_5  \left[ f^g_4f^g_5 (1+f^g_3)-f^g_3(1+f^g_4)(1+f^g_5)\right] (4E-3m)^2 \left[ \frac{\vert M_{45\rightarrow 123}\vert ^2}{s^2}\right]_{s=4mE-3m^2} \;,
\end{eqnarray}
we see that both $R^{gain}_{gcc}$ and $R^{loss}_{gcc}$ contain a same contribution, which is proportional to $f^g_3f^g_4f^g_5$. And $f^g_3$ is the function of total momentum ${\bf P}$.

\section{\label{app:rate_eq_cim}Rate equations for const and isotropic matrix element}
In this appendix we calculate the net rates of the condensation processes for constant and isotropic matrix element, assuming the gluons are in equilibrium. 

\subsection{\label{appB:gc}Rate for $g+c\longleftrightarrow g+g+g$ }
Assuming the gluons in equilibrium, $f^g_i=1/(e^{(E_i-m)/T}-1)$, the term $f^g_3f^g_4f^g_5 (1+f^g_2)-f^g_2(1+f^g_3)(1+f^g_4)(1+f^g_5)$ appeared in Eq.~(\ref{eq:APP_Rgc}) can be simplified 
\begin{eqnarray}
f^g_3f^g_4f^g_5 (1+f^g_2)-f^g_2(1+f^g_3)(1+f^g_4)(1+f^g_5) &=& f^g_3f^g_4f^g_5f^g_2e^{(E-3m)/T}(e^{m/T}-1)\;.
\end{eqnarray}

Thus, Eq.~(\ref{eq:APP_Rgc}) can be rewritten as
\begin{eqnarray}
\label{eq:APPb_Rgc}
R_{gc} &=& \frac{n_c}{3\cdot 2^{11}\cdot \pi^5} (e^{m/T}-1) \frac{\vert M_{345\rightarrow 12}\vert ^2}{m} \int dE_3 dE_4 dE_5 \left[ p_3+p_4-\vert p_3-p_4 \vert\right] \frac{e^{(E-3m)/T}}{e^{(E_3-m)/T}-1} \frac{1}{e^{(E_4-m)/T}-1}\nonumber \\
&& \times  \frac{1}{e^{(E_5-m)/T}-1} \frac{1}{e^{(E_2-m)/T}-1}   \;,
\end{eqnarray} 
where $E_2=E_3+E_4+E_5-m$. Letting $x_4=e^{-E_4/T}$ and $x_5=e^{-E_5/T}$, after a lengthy integration process, we get 
\begin{eqnarray}
R_{gc} &=& \frac{n_c T^4}{3\cdot 2^{10}\cdot \pi^5} (e^{-m/T}-e^{-2m/T}) \frac{\vert M_{345\rightarrow 12} \vert ^2}{m} \Big\{ \int^{e^{-m/T}}_{0} dx_3 \int^{e^{-m/T}}_{x_3} dx_4 \frac{\sqrt{\ln^2 x_4-(m/T)^2}}{(e^{-m/T}-x_3)(e^{-m/T}-x_4)(e^{-m/T}-x_3x_4)}   \nonumber \\  
&& \times \left[ \ln (e^{-m/T}-x_3x_4)-\ln (0) \right]  + e^{m/T}\int^{e^{-m/T}}_{0} dx_3 \frac{\sqrt{\ln^2 x_3-(m/T)^2}}{(e^{-m/T}-x_3)(1-x_3)} \nonumber \\
&& \times \left[ \ln \frac{e^{-m/T}-x^2_3}{e^{-m/T}-x_3} \ln (0) +\frac{1}{2}\ln^2(e^{-m/T}-x^2_3)-\frac{1}{2}\frac{m^2}{T^2} +\int^{x_3}_{0} dx_4 \frac{\ln (e^{-m/T}-x_3x_4)}{e^{-m/T}-x_4} \right]  \Big\}   \;.
\end{eqnarray}

\subsection{\label{appB:ggc}Rate for $g+g+c\longleftrightarrow g+g$ }
Assuming the gluons in equilibrium, $f^g_i=1/(e^{(E_i-m)/T}-1)$, the term $f^g_4f^g_5(1+f^g_2)(1+f^g_3) -f^g_2f^g_3(1+f^g_4)(1+f^g_5)$ appeared in Eq.~(\ref{eq:APP_Rggc}) can be simplified 
\begin{eqnarray}
f^g_4f^g_5(1+f^g_2)(1+f^g_3) -f^g_2f^g_3(1+f^g_4)(1+f^g_5) &=& f^g_4f^g_5f^g_2f^g_3e^{(E_4+E_5-3m)/T}(1-e^{m/T})\;.
\end{eqnarray}
where $E_3=E_4+E_5-E_2-m$. 

Thus, we can rewrite Eq.~(\ref{eq:APP_R_gcc}) and integrate $dE_2$ first, 
\begin{eqnarray}
\label{eq:APPb_Rggc}
R_{ggc} &=& \frac{n_c}{64\pi } (1-e^{m/T}) \frac{\vert M_{45\rightarrow 123}\vert ^2}{m} \int \frac{d^3p_4}{(2\pi)^32E_4}\frac{d^3p_5}{(2\pi)^32E_5} \frac{1}{P} \frac{e^{(E_4+E_5-3m)/T}}{(e^{(E_4-m)/T}-1)(e^{(E_5-m)/T}-1)}    \nonumber \\
&& \times \int^{E^+_2}_{E^-_2} dE_2 \frac{1}{e^{(E_2-m)/T}-1} \frac{1}{e^{(E_4+E_5-E_2-2m)/T}-1} \nonumber \\
&=& \frac{n_cT}{64\pi } (1-e^{m/T}) \frac{\vert M_{45\rightarrow 123}\vert ^2}{m} \int \frac{d^3p_4}{(2\pi)^32E_4}\frac{d^3p_5}{(2\pi)^32E_5} \frac{1}{P} \frac{1}{(e^{(E_4-m)/T}-1)(e^{(E_5-m)/T}-1)(1-e^{(3m-E_4-E_5)/T})}    \nonumber \\
&& \times \ln \frac{\cosh \frac{E-3m+P\sqrt{1-\frac{4m^2}{s-2mE+m^2}}}{2T}-1}{\cosh \frac{E-3m-P\sqrt{1-\frac{4m^2}{s-2mE+m^2}}}{2T}-1} \Theta (E-3m) \Theta ((E-m)^2-{\bf P}^2-4m^2)  \,,
\end{eqnarray}

We integrate the above equation over the solid angles of ${\bf p}_4$ and ${\bf p}_5$, and obtain
\begin{eqnarray}
\label{eq:APPb_Rggc1}
R_{ggc} &=& \frac{n_cT}{64\cdot (2\pi)^5 } (1-e^{m/T}) \frac{\vert M_{45\rightarrow 123}\vert ^2}{m} \int dp_4dp_5 \int d\cos \theta \frac{p^2_4p^2_5}{E_4E_5} \frac{1}{P} \frac{1}{(e^{(E_4-m)/T}-1)(e^{(E_5-m)/T}-1)}    \nonumber \\
&& \times \frac{1}{1-e^{(3m-E_4-E_5)/T}} \ln \frac{\cosh \frac{E-3m+P\sqrt{1-\frac{4m^2}{s-2mE+m^2}}}{2T}-1}{\cosh \frac{E-3m-P\sqrt{1-\frac{4m^2}{s-2mE+m^2}}}{2T}-1} \Theta (E-3m) \Theta ((E-m)^2-{\bf P}^2-4m^2)  \,,
\end{eqnarray}
where $\theta$ is the angle between ${\bf p}_4$ and ${\bf p}_5$.

\subsection{\label{appB:gcc}Rate for $g+c+c\longleftrightarrow g+g$}

Assuming the gluons in equilibrium, $f^g_i=1/(e^{(E_i-m)/T}-1)$, the term $f^g_4f^g_5 (1+f^g_3)-f^g_3(1+f^g_4)(1+f^g_5)$ appeared in Eq.~(\ref{eq:APP_Rgcc}) can be simplified 
\begin{eqnarray}
f^g_4f^g_5 (1+f^g_3)-f^g_3(1+f^g_4)(1+f^g_5) &=& f^g_4f^g_5(1+f^g_3)(1-e^{m/T})\;.
\end{eqnarray}

Thus, Eq.~(\ref{eq:APP_Rgcc}) can be rewritten as
\begin{eqnarray}
\label{eq:APPb_Rgcc}
R_{gcc} &=& \frac{n^2_c}{512\pi^3} (1-e^{m/T}) \frac{\vert M_{45\rightarrow 123}\vert ^2}{m^2} \int dE_4 dE_5  \frac{1}{e^{(E_4-m)/T}-1} \frac{1}{e^{(E_5-m)/T}-1} \frac{e^{(E_3-m)/T}}{e^{(E_3-m)/T}-1}  \Theta(E_4+E_5-3m) \;, 
\end{eqnarray}
where $E_3=E_4+E_5-2m$, $\Theta$ denotes step function. Letting $x_4=e^{-E_4/T}$ and $x_5=e^{-E_5/T}$, after a lengthy integration process, we get 
\begin{eqnarray}
R_{gcc} &=& \frac{n^2_c T^2}{512\pi^3} \frac{\vert M_{45\rightarrow 123} \vert ^2}{m^2} \Big\{ -2 \ln^2 (0) -[11\frac{m}{T}-4ln(e^{m/T}-1)] \ln (0) -4\frac{m^2}{T^2} +6\frac{m}{T}ln(e^{m/T}-1) -\frac{1}{2} ln^2(e^{m/T}-1)  \nonumber \\
&& -\frac{\pi^2}{6} +Li_2(\frac{1}{1-m/T}) +Li_2(e^{m/T}) +Li_2(e^{-m/T}) \Big\}   \;,
\end{eqnarray}
where the last three terms in the curly braces denote the polylogarithm functions $Li_n(z)$ with $n=2$ and $z=1/(1-m/T), e^{m/T}$ and $e^{-m/T}$, respectively.

\end{widetext}

\nocite{*}

\bibliography{References}

\end{document}